# Twist-induced spin splitting and spin-Hall-like effect in antiferromagnetic bilayers

Zhigang Song[1], Xiuying Zhang[2], Julian Klein[3], Jonathan Curtis[1], Frances M. Ross[3], and Prineha Narang[1,*]

1 John A. Paulson School of Engineering and Applied Sciences, Harvard University, Cambridge, MA 02138, USA

2 Department of Physics, National University of Singapore, Science Drive, Singapore, 117551, Singapore

3 Department of Materials Science and Engineering, Massachusetts Institute of Technology, Cambridge, Massachusetts 02139, USA

*corresponding author(s): Prineha Narang (prineha@harvard.seas.edu)

## Abstract

Momentum-resolved spin-polarized bands are a key ingredient in many proposed spintronic devices, but their existence often relies on lattice commensurability or strong spin-orbit coupling. By a large-scale DFT calculation (up to 4212 atoms), we propose a way to realize strongly spin-polarized bands in the absence of these ingredients by twisting monolayers of van der Waals magnetic semiconductor CrSBr. Furthermore, due to the highly anisotropic electronic transport in this material, the twist-induced electronic transport becomes strongly coupled to the spin transport. We show that an in-plane electric field induces a transverse spin current, manifesting a twist-tunable spin-Hall effect in the absence of spin-orbit coupling. Using high-throughput computations, we also identify 231 other material candidates out of a set of 6000 magnetic two-dimensional materials, which satisfy the necessary conditions to realize this behavior, paving the way to widespread application of twist-tunable spin transport.

**Introduction**

Recently there has been a rapid development of spintronic technologies due to the invention and commercial application of magnetic random access memories.[1] A key property for future spintronic applications is the existence of a momentum dependent spin polarization, either due to antiferromagnetic (AFM) exchange interactions (which in turn relies on lattice commensurability), or spin-orbit coupling.[2] Antiferromagnetic materials in particular offer a number of potential benefits due to their comparatively faster picosecond scale spin-dynamics, allowing for more rapid magnetic switching.[2] Furthermore, antiferromagnetic materials are comparatively common and often feature higher critical temperatures than ferromagnets. Reliably exerting control over these antiferromagnetic systems is therefore a key goal for future spintronic applications.

Due to recent experimental breakthroughs, it has become possible to obtain and study magnetism in two-dimensional moiré superlattices, producing a great deal of experimental and theoretical interest.[3-6] Moiré systems have already produced a series of interesting effects in nonmagnetic systems[7-10] including superconductivity and Mott insulating states in twisted bilayer graphene,[11,12] interlayer excitons in heterobilayers of transition metal dichalcogenides (TMDs),[13] and ferroelectricity in twisted hexagonal boron nitride.[14] Twist-angle offers a new and powerful control knob governing the behavior of magnetic materials, with the potential for a wide variety of supercell-resolved magnetic orders, offering possibilities in a new field of moiré spintronics. In order to realize practical spintronic functionalities in moiré superlattices, it is important to be able to generate magnetically ordered phases in a wider array of materials with higher transition temperatures, greater stability, and electronic interfacing, namely momentum-resolved spin polarization. Until now, investigations of spin-polarized electronic states in twisted bilayer materials have been rare in both theory and experiments.

Here, we study twisted layers of van der Waals magnetic semiconductor CrSBr in the method of the large-scale DFT calculation (up to 4212 atoms), and we show that it is ideal for moiré controlled spintronic applications. Due to a combination of large anisotropy in electronic dispersion and strong interlayer antiferromagnetic coupling, relative twist-angle has a large effect on the interlayer magnetic coupling and resulting spin transport. Even in the absence of spin-orbit coupling we find the spin transport becomes locked to the crystalline anisotropy in each layer, leading to an effective layer-polarized spin Hall effect. This interlayer coupling also induces a layer-dependent strain, allowing for the interplay between transport, magnetism, and strain effects. All of these indicate that CrSBr is an ideal candidate for future moiré spintronic devices. We also identify a number of other compounds which have similarly desirable properties which may be well-suited for future studies.

**Results**

 **General theory and material candidates**

We begin by considering the general symmetries of a bilayer system which has intralayer ferromagnetic order and interlayer antiferromagnetism. Consider the action of time-reversal operator $\Theta$ and spatial inversion operator $I$ on the electronic wavefunction $\Psi_\mathbf{k}$ in the moiré bilayer. Under time-reversal we obtain $\Theta\Psi_\mathbf{k} = i\sigma_y\Psi^*_{-\mathbf{k}}$. Subsequently applying spatial inversion we find $I\Theta\Psi_\mathbf{k} = i\sigma_y\tau_x\Psi^*_\mathbf{k}$, where $\tau_x$ is a Pauli matrix acting on layer degree of freedom, such that inversion switches the layer index in the moiré bilayer. We can verify that $(\Theta I)^2 = -1$, and thus even if time-reversal symmetry or inversion symmetry is individually broken, so long as their combination is preserved we will have $(I\Theta)^{-1}H(\mathbf{k})I\Theta = H(\mathbf{k})$. This will imply that eigenvalues come in pairs, with $E^\uparrow(\mathbf{k}) = E^\downarrow(-\mathbf{k})$. We will see that this symmetry is sufficient to destroy the spin-transport effect, but this can be ameliorated by the relative twisting, which disrupts the spatial inversion symmetry.

The second key ingredient besides the symmetry of the twisted bilayer is the electronic anisotropy of each individual monolayer, which can allow for a large momentum-resolved spin polarization in the twisted system without space-time inversion symmetry.[15,16] If the electronic anisotropy of each monolayer is small, the momentum-resolved spin splitting remains weak in the twisted system even in the absence of space-time inversion symmetry. We therefore characterize the electronic anisotropy of a monolayer by considering the effective masses along the *a*- and *b*-directions. Usually, the band structure near the conduction-band minimum (valence-band maximum for hole transport) can be approximated by

$$E_\mathbf{k} = \frac{\hbar^2 k_a^2}{2m_a} + \frac{\hbar^2 k_b^2}{2m_b}$$

, where $m_a$ and $m_b$ are the effective masses in the direction of crystalline *a* and *b* axes of a monolayer respectively. In low symmetry materials, the group velocity in one direction can be much larger than that in another direction $v_{nb} = \frac{\partial E_n}{\partial k_b} \gg v_{na} = \frac{\partial E_n}{\partial k_a}$, with $n$ indexing each distinct band. In general, the transport and optical properties may even approach the one-dimensional limit for large anisotropies.[17] Crucially, in the case of a twisted bilayer, each layer has the a and b axes twisted with respect to the other layer. In this case, many properties including the optical response are then tunable by controlling the twist-angle.[18,19]

To be more specific, we consider the Hamiltonian for the antiferromagnetic twisted bilayer

$$H = \begin{pmatrix} H_t(R_{\vartheta/2}\mathbf{k}) & T_m \\ T_m^\dagger & H_b(R_{-\vartheta/2}\mathbf{k}) \end{pmatrix} \quad (1)$$

written in block-matrix corresponding to each layer, with *t*, *b* indicating the layer index for the top and bottom layer, respectively. $R_{\pm\vartheta/2}$ is the matrix which rotates by an angle of $\pm\vartheta/2$ and acts to transform the lab-frame indexed momentum k in to the local layer coordinate system, with $\vartheta$ the total relative twist angle between the layers. Finally, Tm is the direct interlayer coupling, which is in general dependent on the spin configuration. However, in the absence of spin-orbit coupling this is an identity matrix in spin-space, which is the case we consider here.

Since the interlayer coupling is small in van der Waals materials, the interlayer coupling can be treated by perturbation theory. In absence of spin-orbit coupling we expect that the intralayer magnetic order will imply that there is a large collinear exchange-splitting in each Hamiltonian $H_\ell(k) \sim J(-1)^\ell \sigma_z$ which induces an effective spin-polarization of the Bloch bands in each layer. To lowest order, the effective tunneling between the two layers induced by $T_m$ is then suppressed as $T_{eff} \sim T_m^2/J$ as interlayer hopping will require an electron to enter in the highly excited minority spin band. However, in the presence of a magnetic field a finite effective hopping can be induced either by inducing a finite spin-canting (if the magnetic field is perpendicular to the Néel vector), or a spin-flop transition (if the magnetic field is parallel to the Néel order). This may yield interesting magnetically-tunable spin transport and optical response, though we for the moment focus on the in-plane transport.

In the intrinsic AFM twisted bilayers, the interlayer coupling is weak and the layer and spin remain good quantum numbers. We can see that without twisting, macroscopic spin-dependent physics will cancel between the two layers. This is evident since in this case, the symmetry $\Theta I$ is preserved by the Hamiltonian, even if the dispersion is anisotropic. For instance, the spin-current, defined as $\mathbf{j}_s = (\sigma_z \mathbf{v} + \sigma_z \mathbf{v})/2$ [20,21] (with $\mathbf{v}$ the group velocity of each band) will in this case cancel due to the layers compensating each other once the sum over all momentum states is carried out. After twisting however, the two spin channels will have different group velocity as resolved in the lab frame, as $\mathbf{v}^t = \nabla_k E^t \neq \mathbf{v}^b = \nabla_k E^b$, resulting in a net spin current which does not cancel between the layers. If the electronic anisotropy is large enough, this is particular evident. In this case the material can be treated as an array of 1D wires. Based on a simple Boltzmann transport model, the spin conductivity under an electric field along the θ direction (with respect to the lab-frame x-axis) is

$$\begin{cases} \sigma_x^s = \sigma_0 \sin(\vartheta/2)(\cos(\vartheta) + \cos(-\vartheta)) \\ \sigma_y^s = \sigma_0 \cos(\vartheta/2)(\cos(\vartheta) - \cos(-\vartheta)) \end{cases} \quad (2)$$

where x and y are axes in the lab frame, and $\sigma_0$ is an intrinsic material dependent measure of the in-plane conductivity. $\vartheta$ is the polar angle of the electric field in with respect to the lab frame plane.

According to Eq.2, the spin current is dependent on the electric field direction $\vartheta$, and an interesting case to consider is when the electric field is oriented in the center of the two twisted monolayer axes, such that $\vartheta = 0$. The charge current will then be aligned with the direction of the electric field along *x*, whereas the spin current will be perpendicular to the electric field along *y*. This is similar to the spin-Hall effect, which is usually induced by spin-orbit coupling.[22] In this case, we find a spin-Hall angle of $j_s/j_e = \tan(\vartheta/2)$, making this directly tunable by the twist-angle, unlike in the case where it arises from strong spin-orbit coupling or non-collinear magnetic order.[23] In this case, the transverse spin-current is due to a joint locking of the spin to the layer index, and layer index to the spatial direction. Since the critical temperature for antiferromagnets can be much larger than for ferromagnetic materials, twisted AFM bilayers can serve as an ideal way to generate spin currents at high-temperatures and in the atomically-thin limit, which is highly desirable for spintronic applications.

As discussed above, the spin-splitting and spin-Hall-like effect in twisted bilayers originates from the anisotropy of the corresponding monolayers and the interlayer antiferromagnetic ordering. These are in fact quite generic conditions, and should be realized in many materials not just CrSBr. To this end, we used an open 2D materials database[24] in order to perform high-throughput computation to search for other material candidates which satisfy these general conditions. Specifically, we searched for materials in which monolayers of the material are ferromagnetic with finite net spin-polarization, and in which the energy difference along two different crystalline directions near the Fermi level is larger than 0.5 eV. These two crystalline directions are along two axes for monoclinic, triclinic and orthorhombic systems. For tetragonal systems, these two directions are defined as one axis and the bisector axis. For trigonal systems, the systems are transformed to orthorhombic systems, and two axes perpendicular to each other are characterized. Among more than 6000 2D magnetic materials, we found 231 materials that exhibit a high electronic anisotropy. Amongst these, there are 16 tetragonal, 120 orthorhombic, 5 monoclinic, 43 triclinic and 47 trigonal materials. The crystal structure and their corresponding spin-polarized band structures are shown in the supporting information. Among all, the candidate materials having the chemical formula ABC (A = Cr, Fe or a rare-earth element; B=O, S, Br; C=Cl, Br)[25] exhibit the highest anisotropy, with the typical structure shown in Fig. 1a. Since CrSBr has a high magnetic transition temperature and is air stable, it is also practically favorable for performing experiments,[6,26] although other candidates like CrOCl and FeOCl may also be promising.[27,28] Having established the ubiquity of our predictions, we now proceed on to examine the original case study—CrSBr—in more detail by employing advanced ab initio calculations in order to produce the spin-resolved band-structure in a moiré configuration.

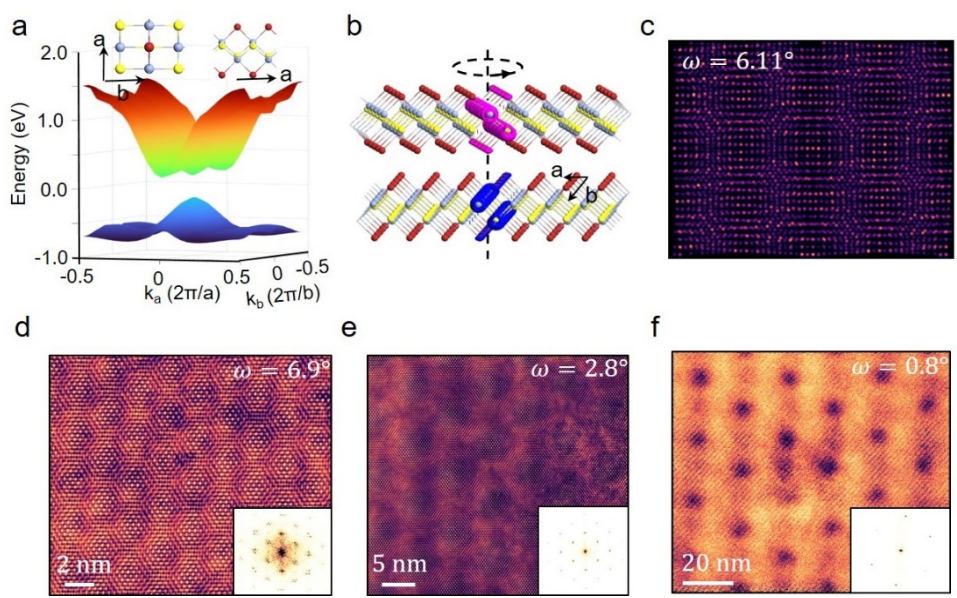

**Figure 1.** Structure of twisted bilayer CrSBr. **a** 3D band structure of monolayer CrSBr. Top and side views of atomic structure are inserted. Gray and yellow, and red balls are Cr, S, and Br atoms, respectively. **b** Atomic structure of twisted bilayer CrSBr and *ab* initio calculated charge density wave near the Fermi level in the range from -0.5eV to 0.5eV. Red and blue color represent the up and down spin channels, respectively. Charge density is cut off at only one unit cell in *a*-direction. **c** Simulated STEM imagine of a bilayer CrSBr with a twist angle of 6.11° based on the DFT relaxed structure containing 4212 atoms. The picture shows a unit cell. **d-f** TEM scanned atomic structures with twist angle 6.9°, 2.8°and 0.8°. Corresponding FFT is inserted

**Electronic structure of twisted bilayer CrSBr**

The *ab* initio calculated band structure of a monolayer CrSBr indeed has a large anisotropy in the electronic dispersion, as shown in Fig. 1a. The dispersion along the a-direction especially in the conduction band is weak and almost flat, whereas it is highly dispersive in the *b*-direction. Indeed, recent work has experimentally demonstrated the strong 1D character of the band structure and optical response of CrSBr, which has also been observed directly in electronic transport.[29] The calculated charge density near the Fermi level shows a series of quasi-one dimensional channels extending along the b-direction (as shown in Fig. 1), despite the slightly larger lattice constant in this direction. This is reflected in the electronic transport, which is greatly enhanced along the *b*-axis as compared to the *a*-axis. The band structure and anisotropy are well described by a symmetry-allowed three-band **k·p** model near the Γ point (see Fig. S1) with Hamiltonian

$$H(\mathbf{k}) = \begin{pmatrix} M_2(\mathbf{k}) & 0 & \frac{\hbar k_y p_y}{m_e} \\ 0 & M_1(\mathbf{k}) & 0 \\ \frac{\hbar k_y p_y}{m_e} & 0 & M_0(\mathbf{k}) \end{pmatrix} \quad (3)$$

where me is the static mass of an electron and $p_y = \langle B_{3g}| \hat{p}_y |B_{1u}\rangle = -1.5\hbar/m_e$ is the optical matrix element between the first and third bands at the Γ point. We have functions $M_2(\mathbf{k}) = E_2 + \hbar^2 k^2/(2m_e)$, $M_1(\mathbf{k}) = E_1 + \frac{\hbar^2 k^2}{2m_e} + \frac{\hbar^2 k_y^2}{m_e^2} \frac{p_y^2}{E_2 - E_0}$ and $M_0(\mathbf{k}) = E_0 + \frac{\hbar^2 k^2}{2m_e} + \frac{\hbar^2 k_y^2}{m_e^2} \frac{p_y^2}{E_0 - E_2}$, with parameters $E_0 = -0.2$ eV, $E_1 = 0.46$ eV and $E_2 = 0.50$ eV the energies at the Γ point, and as in the previous discussions we have neglected spin-orbit coupling. For more details, we refer the reader to the Methods section. Since only one off-diagonal matrix element $\propto p_y$ is nonzero, only photons polarized in the *b*-direction can be absorbed or emitted near the band gap, in line with experiments.[30] Our calculations show that the interband polarization $\eta_\mathbf{k} = \frac{|\langle\psi_{c\mathbf{k}}|\hat{p}_y|\psi_{v\mathbf{k}}\rangle|^2 - |\langle\psi_{c\mathbf{k}}|\hat{p}_x|\psi_{v\mathbf{k}}\rangle|^2}{|\langle\psi_{c\mathbf{k}}|\hat{p}_y|\psi_{v\mathbf{k}}\rangle|^2 + |\langle\psi_{c\mathbf{k}}|\hat{p}_x|\psi_{v\mathbf{k}}\rangle|^2}$ is close to 100% along the y direction for photons near the band edge (see Fig. S2).

We now move on to consider the twisted system. After twisting a bilayer, a moiré superlattice will form.[10,31] We calculate the interlayer coupling energy as a function of the relative layer shift. The interlayer coupling is always antiferromagnetic, unlike in CrI$_3$. Note, due to the antiferromagnetic ordering, there is no net magnetization in real space. The exact atomic structure is obtained by DFT relaxation including collinear spin polarization and van der Waals interactions. As the twist angle decreases, the number of atoms per moiré supercell increases rapidly from 144 to 4212 as the twist angle varies from 90° to 6.11°. The smallest twist angle we calculate for is $\vartheta = 6.11°$. The relaxed structure (see Fig. S2) shows a large atomic reconstruction, resulting in a substantial strain field. This strain field exhibits a chiral vortex in each layer, with opposite chirality in the upper and lower layer. The strain for varying twist angle shows a qualitatively similar pattern but with different magnitude. As the twist angle decreases, both in-plane and out-of-plane components of the atomic displacement field increase. For example, the maximum in-plane displacement is 0.2 Å in the structure with twist angle of 6.11°, and the maximum out-plane displacement is 0.4 Å.

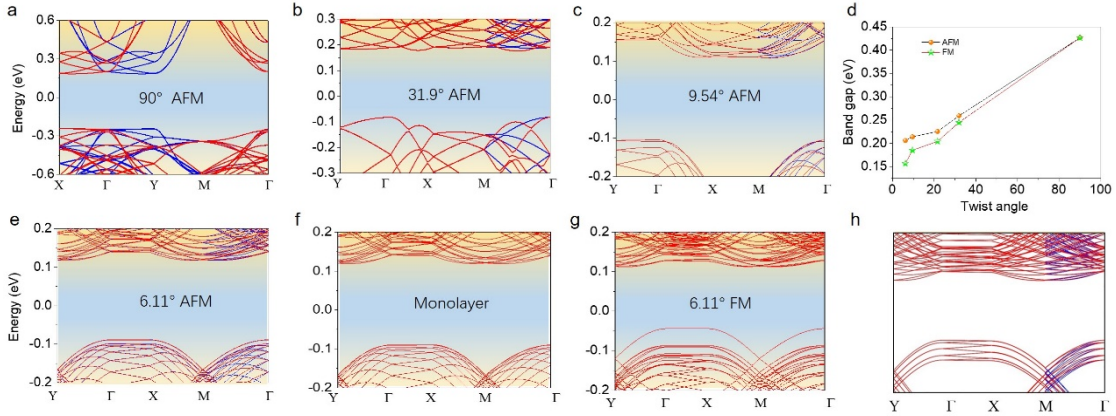

**Figure 2.** Band structures in case of different twist angle and different spin configuration. **a-c** DFT calculated band structures of twisted bilayer CrSBr with twist angles of 90°, 31.9°and 9.54°. The color of the band depicts the majority spin configuration with spin-up in red and spin-down in blue. **d** Band gap as a function of twist angle in ferromagnetic and antiferromagnetic configuration. **e** Band structure of twisted bilayer by 6.11° in antiferromagnetic configure. f Band structure of one monolayer in twisted bilayer by 6.11°. **f** Band structure of twisted bilayer by 6.11° in ferromagnetic configure. **h** Analytical band structure of twisted bilayer by 6.11°

All theory here is based on antiferromagnetic moiré superlattices, but the accessibility and stability of such a superlattice are unknown. Thus, in experiments we created moiré twisted CrSBr samples made from multilayer CrSBr (with thickness ~ 10 nm) and performed high-angle annular dark-field scanning transmission electron microscopy (HAADF-STEM). Images of samples with different twist angles are shown in 1d-f. The pictures show that the nanosheets show (001) surfaces. A series of moiré superlattices with different twist angles are found in the overlap zones stacked by two nanosheets. As the twist angle decreases, we see the moiré wavelength increasing. The simulated STEM image for an angle of 6.11°based on the DFT relaxed structures are in excellent agreement with our experimental images (as shown in Fig. 1c and d). This experimentally confirms the accessibility and stability of CrSBr moiré superlattices and the reliability of DFT calculations for spin-polarized moiré superlattices. Direct experimental confirmation of the predicted spin transport in these superstructures is however beyond the scope of the current apparatus. It however may be possible for future experiments to confirm our predictions using bilayer or multilayer CrSBr, since the effect should be localized to the interface between the two multilayer crystals.

Having both theoretically and experimentally confirmed the viability of this material for generation of moiré superlattices, we now continue by examining the electronic structure of the moiré twisted CrSBr. To obtain the electronic structures, we performed large-scale DFT calculations for twisted bilayer CrSBr with different twist angles. As shown in Fig. 2, the band gap in the antiferromagnetic configuration decreases from 0.42 eV to 0.21 eV as the twist-angle is varied. In the ferromagnetic configuration, the band gap decreases from 0.42 eV to 0.155 eV. While it is possible that the DFT band gap may be underestimated, there is a clear trend of change to the band gap which is more robust. Notably, we find that the interlayer antiferromagnetic configuration is always lower in energy, which may be anticipated from the bulk material properties, which exhibits interlayer antiferromagnetic order. Importantly, the intralayer exchange splitting remains large throughout the twist-angle series.

As the twist angle decrease from 90° to 0°, the momentum-resolved spin-splitting decreases and recovers to zero in the case of zero twist angle due to the recovery of the space-time inversion symmetry, and the spin splitting reaches its maximum for a twist angle of 90°. Comparing the same structures in different spin configurations, we find that ferromagnetic states have a smaller band gap than antiferromagnetic states, and that this energy splitting increases as the twist angle decreases. Interestingly, the band gap is almost the same when the 90° twisted materials switch from an antiferromagnetic state to the ferromagnetic state, possibly indicating an additional decoupling of the layers in this limit.

As shown in Fig. 2e&g, the difference in the band gap between the FM and AFM phases can be as large as 60 meV when the twist angle decreases to 6.11°, whereas the energy difference between the two phases is much smaller and can be switched by a magnetic field of 0.3 T. In normal semiconductors, the Landau level separation at a such a magnetic field is usually several meV or below,[32] indicating that the dominant effect of the magnetic field in this system may indeed be the control of interlayer tunneling as alluded to in our previous discussion. It also may yield an interesting competition between the two phases in the presence of carriers, which may trigger a phase transition once the additional electronic energy overwhelms the weak exchange splitting.

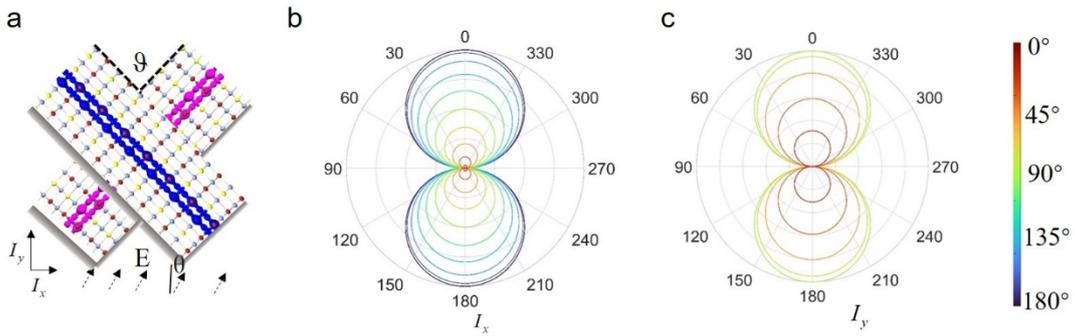

**Figure 3**. Magneto-transport of twisted bilayer CrSBr. **a** Illustration of spin channel. **b-c** Spin conductivity in *y* and *x* direction in case of different twist angle. The conductivity unit is $\sigma_0$. The Fermi level is set 0.1 eV, and spatial angle represents the direction of the applied electric field. Color of curves represents the twist angle $\vartheta$

With a symmetry and group theory analysis, the valence and conduction bands belong to different irreducible representations. The direct coupling between the highest conduction band and lowest valence band is zero. These two bands change independently. Thus, the band gap is sensitive to external perturbations, such as strain. If we maintain the strain induced by the interlayer twist, the band structure of a single layer (as shown in Fig. 2f) is almost identical to that of the antiferromagnetic twisted bilayer, with the difference near the Fermi level less than 1 meV. In contrast, the antiferromagnetic band structure is much different from that of the interlayer ferromagnetic twisted bilayer. The difference between the Fermi level can be as large as 50 meV. This is largely due to the highly suppressed interlayer hopping in the antiferromagnetic case. The interlayer coupling is mediated by the atomic orbitals of deep energy levels and intralayer strain. Since there is no direct interlayer coupling, twist induces a pure periodic strain field without interference from any other interaction. A periodic strain field can have a pseudomagnetic effect on the electronic structure.[33] A clean strain field is desired, but has never realized in any materials.[33-35] After including the strain in our model, we achieved a good accordance between DFT calculation and model. However, the direct interlayer coupling in a ferromagnetic twisted bilayer is nonzero. When an external magnetic field is applied, all spin will tend to be aligned with the magnetic field, and the spin configuration can be switched from AFM ordering to ferromagnetic ordering. The critical magnetic field is only below 0.3 T in CrSBr according to previous experiments.[26] Thus, the spin flipping results in a finite magnetoresistance.

In the antiferromagnetic configuration, strain largely changes the band gap but keeps the quasi-1D characteristic intact. In this case, the upper and lower layers of the twisted bilayer structure have spin up and down, respectively (as shown in Fig.3a). In each layer, the group velocity in the *b*-direction is much larger than the a-direction, $v_{nb} = \partial E_n/\partial k_b \gg v_{na} = \partial E_n/\partial k_a$. Actually, the effective mass ratio along the a- and b-direction for a hole at the first band below the Fermi level is as large as 7.5. For an electron at the first band above the Fermi level, the ratio of effective mass in the *a*- and *b*-direction is as large as 120.5, and for the second band above the Fermi level it is 313.1. In the absence of spin-orbit coupling we have $j_s = s_z v$ in each layer. The spin-polarized current is thus effectively locked to the crystalline axis of each layer. When an electrical field is applied in the plane, a net spin current will be generated vertical to the electronic current. Fig.3b-c show th *a−b* e results of spin current in *x*-and *y*- direction in twisted bilayer CrSBr, when the Fermi level is shifted to 0.1 eV above the conduction band bottom, resulting in electron doping. The spin current is maximum when the electrical field is oriented along the b axes of two layers. In this case, the spin current is not zero but the electron current is zero in the x-direction. The spin current is thus transverse to the electric field, resulting in an anomalous spin Hall effect. As the twist angle increases to 90°, the spin current increases. As the twist angle increases from 90° to 180°, the spin current decreases. Usually, the spin Hall effect is observed in materials with large spin-orbit coupling—here we provide a route towards realizing the spin-Hall current without spin-orbit coupling. This spin current can be observed by measuring the spin torque induced by spin current.[36]

# Methods

## Density functional theory calculation

The calculation of density functional theory is performed on a single-zeta atomic basis set. PBE method is applied to deal with the exchange correlation interaction. The pseudopotential of FHI is applied. DFT-D2 is applied to include the interlayer molecular interaction (Van der Waals interaction). Spin-orbit coupling is neglected, but collinear spin polarization is included. The twisted structures are relaxed until the force on each atom is smaller than the criteria of 0.01 eV/Å.

## Theoretical model

Monolayer CrSBr hold a NO 59 space group. This group has 8 basic symmetries. Four of them are point symmetries $E$, $\sigma(xz)$, $\sigma(yz)$, $C_2(z)$. Other four are combination of point operator and translation, $\tilde{C}_2(x) = \{C_2(x) | \tau = \frac{1}{2}\mathbf{a} + \frac{1}{2}\mathbf{b}\}$, $\tilde{C}_2(y) = \{C_2(y) | \tau = \frac{1}{2}\mathbf{a} + \frac{1}{2}\mathbf{b}\}$, $\tilde{\sigma}(xy) = \{\sigma(xy) | \tau = \frac{1}{2}\mathbf{a} + \frac{1}{2}\mathbf{b}\}$ and $\tilde{i} = \{i | \tau = \frac{1}{2}\mathbf{a} + \frac{1}{2}\mathbf{b}\}$. The character table is list in table

|  | E | $C_2(z)$ | $\tilde{C}_2(y)$ | $\tilde{C}_2(x)$ | $\tilde{i}$ | $\tilde{\sigma}(xy)$ | $\sigma(xz)$ | $\sigma(yz)$ |
|---|---|---|---|---|---|---|---|---|
| $A_g$ | +1 | +1 | +1 | +1 | +1 | +1 | +1 | +1 |
| $B_{1g}$ | +1 | +1 | -1 | -1 | +1 | +1 | -1 | -1 |
| $B_{2g}$ | +1 | -1 | +1 | -1 | +1 | -1 | +1 | -1 |
| $B_{3g}$ | +1 | -1 | -1 | +1 | +1 | -1 | -1 | +1 |
| $A_u$ | +1 | +1 | +1 | +1 | -1 | -1 | -1 | -1 |
| $B_{1u}$ | +1 | +1 | -1 | -1 | -1 | -1 | +1 | +1 |
| $B_{2u}$ | +1 | -1 | +1 | -1 | -1 | +1 | -1 | +1 |
| $B_{3u}$ | +1 | -1 | -1 | +1 | -1 | +1 | +1 | -1 |

As shown in Fig. S1, the direct optical absorption mainly happens at the Γ point. At the Γ point, the valence band transform as $B_{3g}$ irreducible representation, and the first conduction band transform as $A_g$ irreducible representation. The second conduction band transforms as $B_{1u}$ irreducible representation. The dipole operators $\hat{p}_y$ and $\hat{p}_x$ transform as $B_{2u}$ and $B_{3u}$, respectively. $\langle B_{3g} | \hat{p}_x | A_g \rangle = \langle B_{3g} | \hat{p}_x | B_{1u} \rangle = \langle A_g | \hat{p}_x | B_{1u} \rangle = 0$ and $\langle B_{3g} | \hat{p}_y | A_g \rangle = \langle A_g | \hat{p}_y | B_{1u} \rangle = 0$. The only nonzero dipole moment is $\langle B_{3g} | \hat{p}_y | B_{1u} \rangle$. This explains why only *b*-polarized absorption is seen in our PL spectrum. According to the symmetry, we build a three-band **k·p** model. The perturbation is kept up to the second order terms. The obtained Hamiltonian is in written equation 1. Here the basis set is $[|B1u\rangle, |Ag\rangle, |B3g\rangle]^T$. The **k·p** band structure is in good accordance to the DFT calculated one, although there is tiny difference. The **k·p** model thus captures well the character of the monolayer CrSBr materials, especially the anisotropy and optical selection. The atomic displacement field can be fitted by

$$\begin{cases} u_x = -\alpha\left(\sin(2\pi x/L_a) + \sin(2\pi y/L_b)\right)\left[\left|\left(\sin(\pi x/L_a)\sin(\pi x/L_a)\right)\right| + C1\right] \\ u_y = \beta\left(\sin(2\pi x/L_a) - \sin(2\pi y/L_b)\right)\left[\left|\left(\sin(\pi x/L_a)\sin(\pi y/L_b)\right)\right| + C1\right] \\ u_z = \gamma\left(\cos(2\pi x/L_a) + 0.6\cos(2\pi y/L_b)\right) + C2 \end{cases} \quad (4)$$

where $L_a$ and $L_b$ are lattice constants of a moiré supercell in the x- and y-direction, and α, β and γ are parameters of strain field. C1 and C2 are constant. Generally, these parameters depend on twist angle. For example, α = −0.032 Å, β = −0.044 Å, γ = 0.1368 Å in case of twist angle of 6.11°, and C1 = 1.1 and C2 = 0.02. The detailed comparison between DFT calculated values and fitted values is shown in Fig. S3. The strain tensor is $\varepsilon_{ij} = \partial u_i/2\partial r_j + \partial u_j/2\partial r_i$ and $\varepsilon_{ii} = \partial u_i/\partial r_i$. To include the strain, we approximate the **k·p** model in Eq. (3) by a continuum lattice model. Near the Γ point, $k^2 \approx 1 - \cos(k)$ and $k \approx \sin(k)$. The strain correction to the hopping term reads:[37,38] $\delta t_{R',R} \approx (\mathbf{R'-R})\cdot\nabla_r u(\mathbf{r})|_\mathbf{R} \cdot \nabla_{r'-r} t_{r',r}|_{\mathbf{R'-R}} = -g t_{R',R}|\mathbf{R'-R}|^{-2}(\mathbf{R'-R})\cdot\nabla_r u(\mathbf{r})|_\mathbf{R}\cdot(\mathbf{R'-R})$, were **R'** and **R** are the lattice vectors in monolayer materials. **r/r'** is the Wannier orbital position measured from a center of a unit cell. We have adopted exponentially decaying hopping integrals $t_{\mathbf{r',r}} = t_{\mathbf{R',R}} e^{-g(|\mathbf{r'-r}|-|\mathbf{R'-R}|)/|\mathbf{R'-R}|}$ where g is the Grunisen parameter of order unity. To the linear order in strain, it consequently contributes to the Hamiltonian as

$$-g\begin{pmatrix} (\varepsilon_{yz}+\varepsilon_{yy}+\varepsilon_{zz})(t_2\cos(k_y)+t_x\cos(k_x)) & 0 & (\varepsilon_{yz}+\varepsilon_{yy}+\varepsilon_{zz})it_3\sin(k_y) \\ 0 & (\varepsilon_{yz}+\varepsilon_{yy}+\varepsilon_{zz})(t_1\cos(k_y)+t_x\cos(k_x)) & 0 \\ -(\varepsilon_{yz}+\varepsilon_{yy}+\varepsilon_{zz})it_3\sin(k_y) & 0 & (\varepsilon_{yz}+\varepsilon_{yy}+\varepsilon_{zz})(t_0\cos(k_y)+t_x\cos(k_x)) \end{pmatrix}$$

$t_0 = 6.8$ eV, $t_1 = -7.4$ eV, $t_2 = -0.5$ eV, $t_3 = -1.5$ eV, and $t_x = -0.0763$ eV. Besides, strain may break some symmetries. The vanished terms can recover after the layer stacking and reconstruction is included. In case of CrSBr,

$$\begin{pmatrix} 0 & it_4\sin(k_y/2)\cos(k_x/2) & 0 \\ -it_4\sin(k_y/2)\cos(k_x/2) & 0 & 0 \\ 0 & 0 & 0 \end{pmatrix} \quad (5)$$

This term is actually the coupling between $|B_{1u}\rangle$ and $|A_g\rangle$ inside two sublattice in a unit cell. All the parameters can be fitted by the DFT calculations. The electron Gruneisen parameter for CrSBr is obtained from a careful comparison of first-principles calculations of the electronic structure of strained CrSBr and the model presented in this work or by estimation of the electron-phonon coupling measured in experiments. After the numerical evaluation, we obtained g=0.0 and $t_4 = 0.09 t_0$ for the twist angle of 6.11°. The interaction induced by symmetry breaking is a short-wave interaction inside a unit cell. The strain effect is a long-wave interaction beyond a unit cell. The interaction induced by symmetry breaking is much stronger than the long-range strain.

**Sample fabrication and characterization of twisted CrSBr**

Bulk CrSBr was grown by chemical vapor transport.[6] CrSBr flakes were mechanically exfoliated and moiré twisted structures are fabricated using dry visco-elastic transfer. The precision of the transfer is 1°. STEM imaging was performed with a probe-corrected Thermo

Fisher Scientific Themis Z G3 S/TEM operated at 200 kV with the probe convergence semi-angle of 19 mrad. The probe size of the aberration-corrected electron beam is sub-Angstrom (0.6 Å). We used a collection semi-angle of 63-200 mrad for STEM-HAADF imaging. The data were collected using the Velox software (Thermo Fisher) with a typical frame size of 1024x1024 pixel and adwell time of 500 ns/pixel.

**Acknowledgements**
This work is partially supported by the Office of Naval Research (ONR) MURI Program under grant number ONR N00014-21-1-2537 as well as the by the U.S. Department of Energy, Office of Science, Basic Energy Sciences (BES), Materials Sciences and Engineering Division under FWP ERKCK47 'Understanding and Controlling Entangled and Correlated Quantum States in Confined Solid-state Systems Created via Atomic Scale Manipulation' (spin qubits), and by the the Army Research Office MURI (Ab Initio Solid-State Quantum Materials) grant number W911NF-18-1-0431 (computational methods for low dimensional materials). P.N. acknowledges support as a Moore Inventor Fellow through Grant No. GBMF8048 and gratefully acknowledges support from the Gordon and Betty Moore Foundation as well as support from a NSF CAREER Award under Grant No. NSF-ECCS-1944085. J.K. acknowledges support by the Alexander from Humboldt foundation. F.M.R. acknowledges the funding from the U.S. Department of Energy, Office of Basic Energy Sciences, Division of Materials Sciences and Engineering under Award DE-SC0019336 for STEM characterization

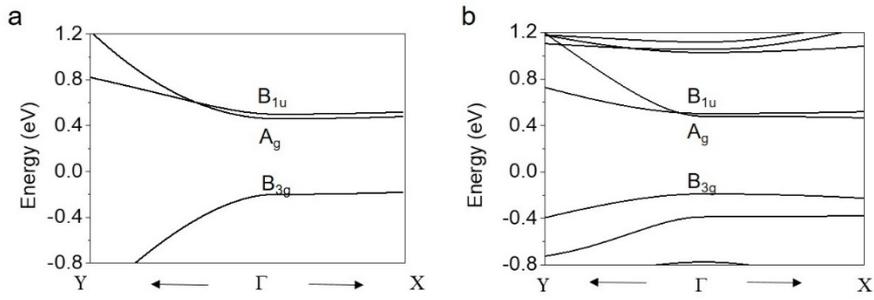

**Figure S1**. Band structure of monolayer CrSBr. **a** Band structure near the Γ point calculated in the method of **k·p**. **b** Band structure near the Γ point calculated by DFT.

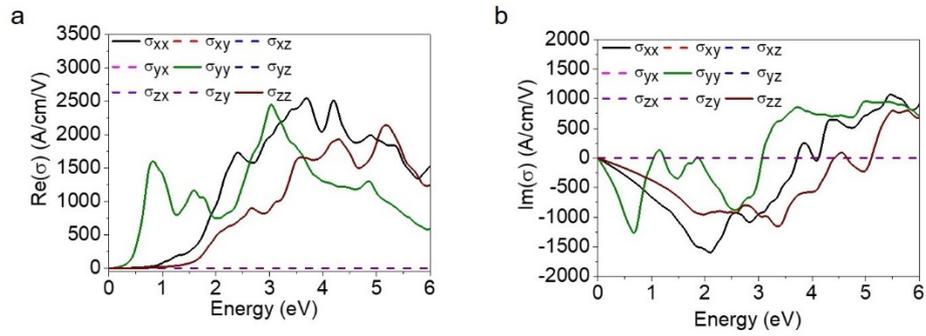

**Figure S2** Optical conductivity of monolayer CrSBr. **a** Real part of optical conductivity. **b** Imaginary part of the optical conductivity. **c** Linear polarization η between the first conduction band and the first valence band near the Fermi level.

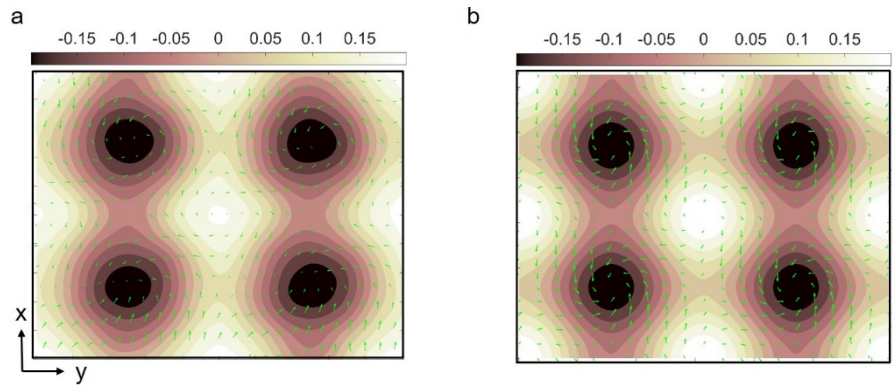

**Figure S3**. Atomic displacement of AFM twisted bilayer CrSBr. **a** DFT calculated atomic displacement of one layer of the AFM twisted bilayer CrSBr with twisted angle of 6.11°. **b** Fitted atomic displacement. The color map represents the out-plane atomic displacement, and the arrows represent the in-plane atomic displacement. The unit is Å.